\documentclass[reprint,amsmath,amssymb,aps,floatfix,longbibliography]{revtex4-2}

\usepackage{graphicx}
\usepackage{dcolumn}
\usepackage{bm}
\usepackage[export]{adjustbox}
\usepackage[version=4]{mhchem}
\usepackage{multirow}
\usepackage{xcolor}
\usepackage{ragged2e}
\usepackage{perpage}
\usepackage{xurl}

\MakePerPage{footnote}

\usepackage[colorlinks=true,
            linkcolor=blue,
            citecolor=blue,
            urlcolor=blue]{hyperref}

\begin{document}

\title{Chemical tuning of magnetic ordering and cryogenic magnetocaloric response in \\ zircon-type Gd$_{1-x}$Er$_x$VO$_4$}

\author{Ming Zeng$^{1}$}
\email{mingzeng@um.edu.mo}

\author{Muqing Su$^{1}$}

\author{Liang Ming$^{1}$}

\author{Xiaolong Yang$^{1}$}

\author{Wang Chen$^{2,1}$}

\author{Lingwei Li$^{2}$}

\author{Hai-Feng Li$^{3}$}
\email{haifengli@um.edu.mo}

\affiliation{$^{1}$Institute of Applied Physics and Materials Engineering, University of Macau, Taipa, Macao S.A.R. 999078, China}
\affiliation{$^{2}$School of Electronics and Information, Hangzhou Dianzi University, Hangzhou 310018, China}
\affiliation{$^{3}$Macao Centre for Research and Development in Advanced Materials, Institute of Applied Physics and Materials Engineering, University of Macau, Taipa, Macao S.A.R. 999078, China}

\date{\today}

\begin{abstract}

Chemical substitution offers an effective route to tune magnetic ordering and magnetocaloric performance in rare-earth oxides for cryogenic refrigeration. Here we investigate the structural evolution, magnetic properties, and magnetocaloric effect of polycrystalline zircon-type Gd$_{1-x}$Er$_x$VO$_4$ ($x = 0$, $0.1$, $0.25$, $0.5$, and $0.75$). Powder x-ray diffraction confirms that all samples crystallize in the tetragonal zircon structure without detectable impurity phases. Substitution of Gd$^{3+}$ by the smaller Er$^{3+}$ ion produces a systematic lattice contraction and modifies the magnetic behavior of the rare-earth sublattice. In particular, the magnetic ordering temperature is suppressed from 3.65(2)~K in GdVO$_4$ to 2.76(2)~K in Gd$_{0.9}$Er$_{0.1}$VO$_4$, accompanied by a weakening of the spin-flop-like field-induced anomaly observed in the parent compound. A low Er concentration correspondingly improves the low-temperature magnetocaloric performance, with Gd$_{0.9}$Er$_{0.1}$VO$_4$ exhibiting a maximum magnetic entropy change of 45.1~$\mathrm{J\,kg^{-1}\,K^{-1}}$ for $\mu_0 \Delta H = 7$~T. These results demonstrate that weak Er substitution effectively tunes the competition among exchange interactions, dipolar coupling, and magnetic anisotropy, optimizing the balance between magnetic ordering and available spin entropy in zircon-type rare-earth vanadates, which is crucial for developing efficient cryogenic refrigeration materials.

\end{abstract}

\maketitle

\section{Introduction}

Low-temperature refrigeration, particularly in the liquid-helium regime, is essential for a wide range of modern scientific and technological applications, including superconducting devices, quantum technologies, and other cryogenic instrumentation \cite{NatureAstronomy2021Driver,zhu2024Nature,huangNature2024,doiron2003Nature}. Because helium is a limited and nonrenewable resource, improving the efficiency of cryogenic cooling and developing alternative solid-state refrigeration strategies are of considerable current interest \cite{hu2025review,rosen2023impending}. Among the available approaches, magnetic refrigeration based on the magnetocaloric effect (MCE) is especially attractive because it provides an efficient, compact, and gas-free route to low-temperature cooling \cite{manosa2020solid,lloveras2021advances,Li2023Science,Du2025,zheng2023colossal,zeng2024AS,zeng2025PST}.

\begin{figure*}[htbp]
\centering
\includegraphics[width=0.78\textwidth]{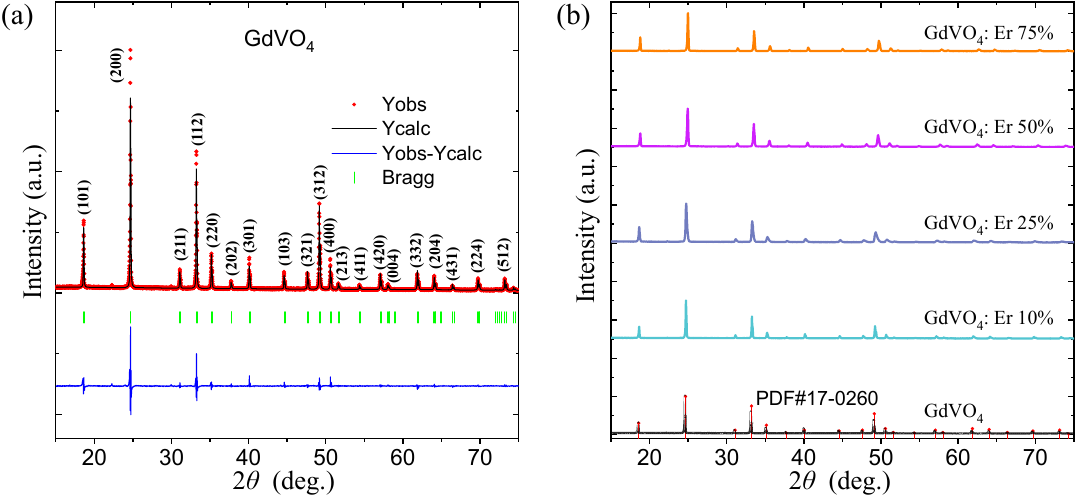}
\caption{
Powder x-ray diffraction analysis of Gd$_{1-x}$Er$_x$VO$_4$ measured at $\sim$ 298 K, with main diffraction peaks indexed by their corresponding (hkl) reflections. (a) Rietveld refinement of the powder x-ray diffraction pattern of GdVO$_4$. Solid red circles represent the experimental data, the black line the calculated profile, the vertical ticks the Bragg reflection positions, and the blue curve the difference between observed and calculated intensities. The refinement confirms the zircon-type tetragonal structure with space group $I4_1/amd$.
(b) Powder x-ray diffraction patterns of Gd$_{1-x}$Er$_x$VO$_4$ ($x = 0$, $0.1$, $0.25$, $0.5$, $0.75$). The systematic shift of the diffraction peaks toward higher $2\theta$ with increasing Er concentration indicates lattice contraction (Table~\ref{tab:table1}) associated with the smaller ionic radius of Er$^{3+}$ relative to Gd$^{3+}$.
}
\label{fig:Figure1}
\end{figure*}

The MCE has been studied for over a century, starting with the early work of Weiss and the subsequent demonstration of adiabatic demagnetization refrigeration by Giauque and MacDougall \cite{weiss1921phenomene,giauque1933attainment}. In recent years, strong cryogenic MCEs have been reported in a broad range of materials, including rare-earth halides, hydroxides, oxysalts, molecular clusters, geometrically frustrated oxides, and other gadolinium-based refrigerants operating down to the sub-liquid-helium regime \cite{NH4GdF4,EuCl2,GdOHF2,LLW2026NC,CHEN2025STRUCTURAL,SJ2025,LB2023,LLuis2024,XJS2024Nature,Shu2026,ZHANG2024119946,LIN202451269,ZHANG2025121033}. These studies collectively indicate that an effective cryogenic magnetic refrigerant should combine a large magnetic entropy with a sufficiently low magnetic ordering temperature, so that the available spin entropy remains accessible within the relevant operating window.

Rare-earth ions with large magnetic moments are natural candidates for cryogenic magnetic refrigeration. In particular, Gd$^{3+}$ is especially attractive because its spin-only ground state, characterized by $S = J = 7/2$ and $L = 0$, results in an almost isotropic magnetic character and enables a substantial magnetic entropy change. However, in many Gd-based compounds, the onset of magnetic ordering at a few kelvin partially consumes the available entropy and correspondingly limits the refrigeration performance at still lower temperatures. Consequently, large magnetocaloric effects in this ultra-low temperature range are typically achieved through strong magnetization changes that occur well away from any antiferromagnetic phase transitions, where the paramagnetic entropy can be fully utilized. A central materials-design challenge is therefore to suppress the ordering temperature without excessively sacrificing the large rare-earth-derived entropy. The magnetic anisotropy of the two rare-earth ions, Gd$^{3+}$ and Er$^{3+}$, primarily arises from differences in their 4\emph{f} electron configurations and total angular momentum quantum number $J$. Gd$^{3+}$ has $J= 7/2$  and a half-filled 4\emph{f} shell (with each spin either up or down), whereas Er$^{3+}$ has $J=15/2$ and 4$f^{11}$ configuration, equivalent to 4$f^{3}$ holes. As a result, Gd$^{3+}$ exhibits extremely weak magnetic anisotropy, while Er$^{3+}$ displays strong magnetic anisotropy. As the concentration of Er$^{3+}$ increases, the relative proportion of Gd$^{3+}$ magnetic ions occupying lattice sites decreases. This dilutes the indirect or direct Gd$^{3+}$--Gd$^{3+}$ superexchange coupling, which is further disturbed by the resulting disorder.

\begin{table*}[htbp]
\caption{\label{tab:table1}
Refined crystallographic parameters of Gd$_{1-x}$Er$_x$VO$_4$ ($x = 0$, $0.1$, $0.25$, $0.5$, $0.75$) obtained from Rietveld refinement of powder x-ray diffraction data. The compounds crystallize in the zircon-type tetragonal structure with space group $I4_1/amd$ (No. 141). Gd occupies the Wyckoff $4a$ site $(0, 0, 0)$, V the $4b$ site $(0, 0, 1/2)$, and O the $16h$ site $(0, y, z)$. The lattice is characterized by the parameters $a(=b)$ and $c$, with $\alpha = \beta = \gamma = 90^\circ$.}
\begin{ruledtabular}
\begin{tabular}{ccccccc}
 & $x=0$ & $x=0.1$ & $x=0.25$ & $x=0.5$ & $x=0.75$ & ErVO$_4$ \\
\hline
$a$ (\AA) & 7.2097(1) & 7.1996(1) & 7.1842(2) & 7.1471(1) & 7.1192(1) & 7.096 \\
$c$ (\AA) & 6.3462(1) & 6.3400(1) & 6.3289(2) & 6.3058(1) & 6.2878(1) & 6.273 \\
$V$ (\AA$^3$) & 329.873(9) & 328.629(8) & 326.651(1) & 322.105(2) & 318.689(5) & 315.8 \\
\hline
O ($y$) & 0.18318 & 0.18004 & 0.18426 & 0.18200 & 0.18250 & -- \\
O ($z$) & 0.32992 & 0.33015 & 0.32520 & 0.32780 & 0.32463 & -- \\
\hline
$R_{wp}$ (\%) & 25.6 & 14.3 & 23.0 & 15.9 & 14.9 & -- \\
$R_p$ (\%) & 22.0 & 14.2 & 25.3 & 17.1 & 14.8 & -- \\
$\chi^2$ & 2.26 & 1.86 & 4.79 & 2.06 & 1.51 & -- \\
\end{tabular}
\end{ruledtabular}
\end{table*}

\begin{figure}[htbp]
\centering
\includegraphics[width=0.45\textwidth]{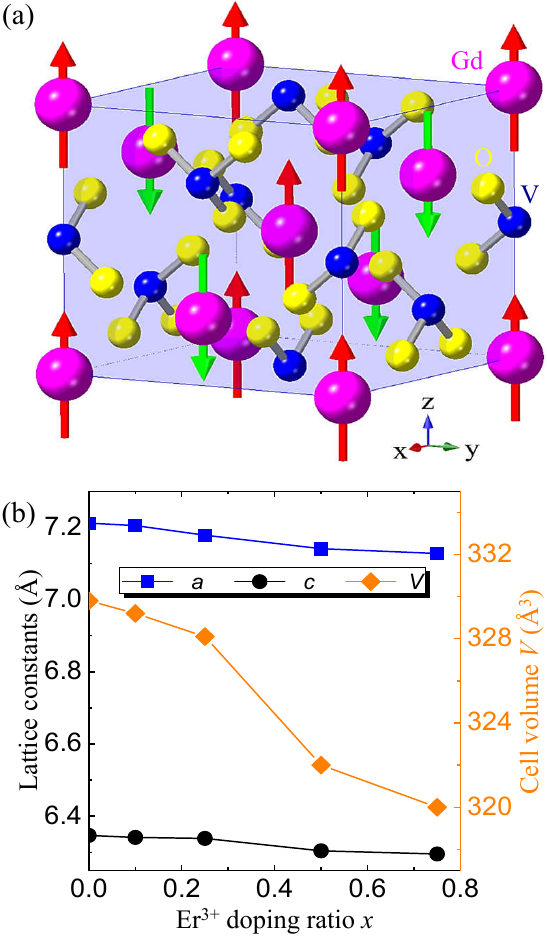}
\caption{
Tetragonal crystal structure and lattice dependence on the Er$^{3+}$ doping ratio in Gd$_{1-x}$Er$_x$VO$_4$.
(a) Crystal and magnetic structures of GdVO$_4$. Purple, yellow, and blue spheres represent Gd, O, and V atoms, respectively. Arrows indicate the orientations of the Gd magnetic moments, illustrating antiferromagnetic coupling between nearest-neighbor rare-earth ions.
(b) Composition dependence of the lattice parameters $a(=b)$ and $c$, and the unit-cell volume $V$ for Gd$_{1-x}$Er$_x$VO$_4$. The monotonic decrease reflects the smaller ionic radius of Er$^{3+}$ relative to Gd$^{3+}$.
}
\label{fig:Figure2}
\end{figure}

\begin{figure*}[htbp]
\centering
\includegraphics[width=0.78\textwidth]{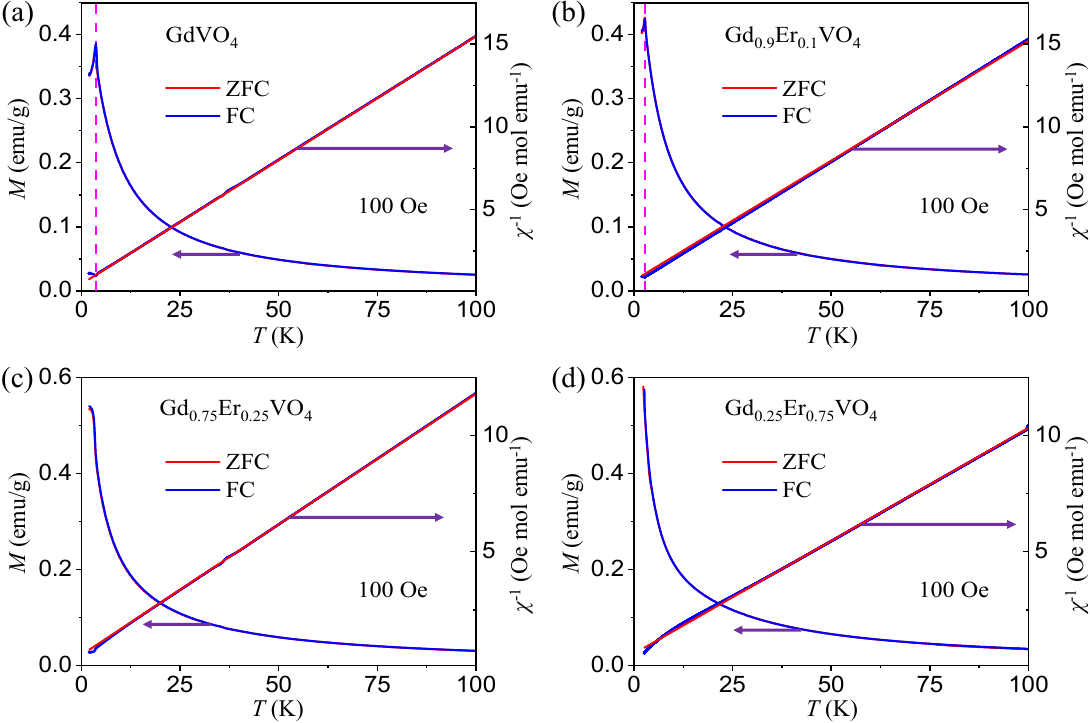}
\caption{
$M(T)$ and corresponding $\chi^{-1}(T)$ curves of Gd$_{1-x}$Er$_x$VO$_4$ measured under an applied field of 100~Oe using ZFC and FC protocols. (a) $x=0$, (b) $x=0.1$, (c) $x=0.25$, and (d) $x=0.75$. GdVO$_4$ exhibits an antiferromagnetic transition at $T_{\mathrm{N}} = 3.65$~K. Er substitution suppresses the magnetic ordering temperature, consistent with a modification of the magnetic interactions within the rare-earth sublattice.
}
\label{fig:Figure3}
\end{figure*}

\begin{table}[htbp]
\caption{\label{tab:table2}
Antiferromagnetic transition temperature $T_{\mathrm{N}}$ and Curie--Weiss temperature $\theta_{\mathrm{CW}}$ obtained from susceptibility analysis of Gd$_{1-x}$Er$_x$VO$_4$. Negative $\theta_{\mathrm{CW}}$ values indicate dominant antiferromagnetic interactions.}
\begin{ruledtabular}
\begin{tabular}{ccc}
$x$ & $T_{\mathrm{N}}$ (K) & $\theta_{\mathrm{CW}}$ (K) \\
\hline
0    & 3.65(2) & -3.42(4) \\
0.1  & 2.76(2) & -4.98(2) \\
0.25 & $<2$    & -4.95(3) \\
0.5  & $<2$    & -6.40(4) \\
0.75 & $<2$    & -6.29(2) \\
\end{tabular}
\end{ruledtabular}
\end{table}

Chemical substitution provides a direct route to tune this balance. In rare-earth compounds, partial substitution can modify interionic spacing, exchange pathways, dipolar interactions, and magnetic anisotropy, thereby reshaping both the zero-field magnetic ground state and the field-induced magnetic response. More generally, the strong sensitivity of rare-earth oxides to structural perturbations has also been highlighted by pressure-tuned spin switching in compensated ferrimagnets, underscoring how lattice modification can alter the balance of competing magnetic interactions and drive spin reconfiguration \cite{PhysRevB.103.054423}. This strategy is particularly appealing in zircon-type rare-earth vanadates, $R$VO$_4$, where nonmagnetic V$^{5+}$ ions separate the rare-earth moments and the magnetic properties are governed primarily by the rare-earth sublattice. GdVO$_4$ crystallizes in the tetragonal $I4_1/amd$ structure and undergoes antiferromagnetic ordering at low temperature, making it a suitable platform for examining how rare-earth substitution perturbs magnetic interactions in a weakly ordered cryogenic magnetocaloric system.

Among the possible substituents, Er$^{3+}$ is of particular interest. Its smaller ionic radius of 1.00~\AA\ for eightfold coordination, compared with 1.05~\AA\ for Gd$^{3+}$ in the same coordination environment, introduces chemical pressure through lattice contraction. In addition, the strong spin--orbit coupling of Er$^{3+}$, characterized by the quantum numbers $S = 3/2$, $L = 6$, and $J = 15/2$, together with its pronounced crystal-field sensitivity, can enhance magnetic anisotropy. More broadly, recent work on mixed rare-earth oxides has shown that chemical substitution can shift low-temperature magnetic transitions, modify field-induced magnetic behavior, and enhance magnetocaloric performance through changes in magnetic anisotropy and spin reconfiguration \cite{D4CP02000F}. Er substitution is therefore expected to influence not only the magnetic ordering temperature but also the character of the spin-flop-like field-induced anomaly and, in turn, the low-temperature MCE. Despite previous studies on rare-earth vanadates, the extent to which weak Er substitution tunes the competition among exchange interactions, dipolar coupling, and magnetic anisotropy in GdVO$_4$ remains insufficiently clarified.

In this work, we seek to elucidate the structural evolution, magnetic properties, and magnetocaloric effect of polycrystalline Gd$_{1-x}$Er$_x$VO$_4$ for values of $x = 0$, $0.1$, $0.25$, $0.5$, and $0.75$, thereby laying the foundation for the development of advanced magnetic refrigerants. By combining structural refinement, temperature- and field-dependent magnetization measurements, and magnetic-entropy analysis, we show that a low Er concentration suppresses the magnetic ordering temperature, weakens the spin-flop-like field-induced anomaly observed in GdVO$_4$, and enhances the cryogenic magnetocaloric performance in the low-temperature regime. These results establish weak rare-earth substitution as an effective route to control the interplay between magnetic ordering and available spin entropy in zircon-type rare-earth vanadates, and they provide useful guidance for designing low-temperature magnetic refrigerants based on rare-earth oxide platforms.

\section{Experimental Methods}

Polycrystalline Gd$_{1-x}$Er$_x$VO$_4$ ($x = 0$, $0.1$, $0.25$, $0.5$, and $0.75$) samples were prepared by a conventional solid-state reaction method. High-purity Gd$_2$O$_3$ (99.99\%), Er$_2$O$_3$ (99.99\%), and V$_2$O$_5$ (99.99\%) powders were used as starting materials to ensure the integrity of the synthesized samples. Stoichiometric amounts of the precursors were thoroughly mixed and ball milled (FRITSCH PULVERISETTE 0) for 120 min to improve homogeneity. The resulting mixtures were pressed into cylindrical pellets at 20 MPa and 200~$^\circ$C for 30 minutes, calcined in air at 800~$^\circ$C for 10 hours, reground and repelletized, and then sintered at 1200~$^\circ$C for 24 hours in air. The samples were finally cooled naturally to room temperature.

Powder x-ray diffraction measurements were conducted using a Rigaku SmartLab 9 kW diffractometer over the $2\theta$ range of $10^\circ$--$80^\circ$ with a step size of $0.02^\circ$. The diffraction data were analyzed by Rietveld refinement using the FullProf Suite \cite{fullprof,li2008synthesis,li2007correlation}. The refined structural parameters were used to track the compositional evolution of the crystal structure across the series.

Magnetic measurements were carried out using the vibrating sample magnetometer option of a Physical Property Measurement System (PPMS DynaCool, Quantum Design). Temperature-dependent magnetization was measured between 2 and 100~K under magnetic fields up to 7~T. Zero-field-cooled (ZFC) and field-cooled (FC) magnetization curves were recorded under an applied field of 100~Oe. Isothermal magnetization curves $M(\mu_0H)$ were collected in the temperature range from 2 to 30~K. Unless otherwise noted, only representative compositions are shown in selected figures for clarity, although all synthesized samples, including the $x = 0.5$ composition, were measured and analyzed.

\section{Results and Discussion}

The key question in Gd$_{1-x}$Er$_x$VO$_4$ is how Er substitution alters the magnetic energy scales that dictate low-temperature ordering and field-induced entropy release. We therefore first examine the structural evolution of the zircon lattice and the local environment of the rare-earth ions (Figs.~\ref{fig:Figure1} and \ref{fig:Figure2}), using the magnetic structure of GdVO$_4$ reported in Ref.~\cite{Palacios2018} as a reference. We then relate these structural changes to the suppression of magnetic ordering revealed by temperature-dependent magnetization and Curie--Weiss analysis (Fig.~\ref{fig:Figure3}). The evolution of the spin-flop-like field-induced anomaly is subsequently analyzed through isothermal magnetization, $dM/d(\mu_0H)$, and Arrott plots (Figs.~\ref{fig:Figure4} and \ref{fig:Figure5}). Finally, the consequences for the magnetocaloric performance are evaluated from the magnetic-entropy change and refrigeration metrics (Figs.~\ref{fig:Figure6}--\ref{fig:Figure8}).

\subsection{Structural evolution and rare-earth local environment}

Because the magnetic interactions in Gd$_{1-x}$Er$_x$VO$_4$ are controlled by the geometry of the rare-earth--oxygen network, the starting point is to establish how the zircon lattice evolves with Er substitution. In this system, even modest changes in lattice dimensions and local coordination are expected to influence the competition among exchange interactions, dipolar coupling, and crystal-field effects.

Figure~\ref{fig:Figure1} establishes that all investigated compositions retain the zircon-type tetragonal structure. The Rietveld refinement for GdVO$_4$ and the composition-dependent powder x-ray diffraction patterns show no detectable impurity phases, indicating that Er is incorporated into the host lattice without changing the crystallographic framework. The refinement confirms that GdVO$_4$ crystallizes in the tetragonal zircon structure with space group $I4_1/amd$ \cite{chakoumakos1994crystal,szczeszak2014structural,dey2017cryogenic}.

With increasing Er concentration, the diffraction peaks shift systematically toward higher $2\theta$. The main peak near $2\theta \approx 25.10^\circ$ corresponds to the (200) reflection. Because Er$^{3+}$ has a smaller ionic radius than Gd$^{3+}$, substitution leads to a gradual lattice contraction, as reflected in the shifts of the (200), (112), and (312) reflections.

The crystal and magnetic structures of GdVO$_4$ are illustrated in Fig.~\ref{fig:Figure2}(a). Each rare-earth ion is surrounded by eight oxygen atoms, forming a distorted triangular dodecahedron, whereas V$^{5+}$ ions occupy tetrahedral VO$_4$ units. Magnetic interactions occur primarily between the rare-earth ions, which are antiferromagnetically coupled to their nearest neighbors.

The compositional dependence of the lattice parameters and unit-cell volume is summarized in Fig.~\ref{fig:Figure2}(b) and Table~\ref{tab:table1}, where the data for ErVO$_4$ are taken from Ref.~\cite{chakoumakos1994crystal}. Both $a(=b)$ and $c$ decrease gradually with increasing Er content, leading to a monotonic reduction in the unit-cell volume. This lattice contraction modifies the RE--O bond geometry and, correspondingly, the local crystal-field environment of the rare-earth ions. Since the magnetic interactions are mediated through these structural units, the observed structural evolution is expected to influence both the magnetic ordering and the field-induced magnetic anomaly discussed below.

\subsection{Suppression of magnetic ordering and Curie--Weiss behavior}

The most direct consequence of Er substitution is a change in the low-temperature magnetic ordering. Temperature-dependent magnetization therefore provides a first probe of how the lattice contraction and modified local environment affect the effective interactions within the rare-earth sublattice.

Figure~\ref{fig:Figure3}(a)--\ref{fig:Figure3}(d) shows the ZFC and FC magnetization curves measured under an applied field of 100~Oe. For all samples, the magnetization increases gradually upon cooling, indicating predominantly paramagnetic behavior over most of the measured temperature range.

A clear trend emerges with increasing Er content. GdVO$_4$ shows an antiferromagnetic transition at $T_{\mathrm{N}} = 3.65(2)$~K, whereas partial substitution by Er suppresses the magnetic ordering temperature, shifting it to 2.76(2)~K in Gd$_{0.9}$Er$_{0.1}$VO$_4$ and likely below 2~K for higher Er concentrations. In contrast, ErVO$_4$ orders antiferromagnetically only at much lower temperatures (around 0.4~K) \cite{AFMErVO4}.

To quantify how the dominant magnetic interactions evolve across the series, the high-temperature susceptibility was analyzed using the Curie--Weiss law [Eq.~(\ref{eq:curie-weiss})]:
\begin{equation}
\chi^{-1} = \frac{T - \theta_{\mathrm{CW}}}{C},
\label{eq:curie-weiss}
\end{equation}
where $C$ is the Curie constant and $\theta_{\mathrm{CW}}$ is the Curie--Weiss temperature. Here
\begin{equation}
C = \frac{N_{\mathrm{A}} \mu_{\mathrm{eff}}^2 \mu_{\mathrm{B}}^2}{3k_{\mathrm{B}}},
\label{eq:mueff}
\end{equation}
where $N_{\mathrm{A}}$ is Avogadro's number, $\mu_{\mathrm{B}}$ is the Bohr magneton, and $k_{\mathrm{B}}$ is the Boltzmann constant. The extracted Curie--Weiss temperatures are all negative (Table~\ref{tab:table2}), indicating dominant antiferromagnetic interactions throughout the series. The magnitude of $\theta_{\mathrm{CW}}$ increases slightly with Er substitution, suggesting that the effective magnetic interactions within the rare-earth sublattice are modified as the composition changes.

The effective magnetic moments obtained from the Curie--Weiss analysis increase with increasing Er concentration, consistent with the growing contribution of Er$^{3+}$ ions. Minor deviations from simple compositional interpolation may reflect the influence of Gd--Er interactions and crystal-field effects. As the Er concentration increases, the strength of Gd--Gd exchange interactions generally shows a downward trend, while the coercive force varies non-monotonically. The temperature- and field-dependent phase transition mechanism shifts from exchange-dominated to anisotropy-dominated. This ultimately manifests as a nonlinear decrease in both the Curie temperature and $T_{\mathrm{N}}$.

\subsection{Field-induced magnetic response and evolution of the spin-flop-like field-induced anomaly}

Beyond the zero-field magnetic ordering, Er substitution also reshapes how the rare-earth moments respond to an applied field. The field dependence of the magnetization therefore provides insight into how the competition among exchange interactions, magnetic anisotropy, dipolar coupling, and Zeeman energy evolves across the series.

\begin{figure*}[htbp]
\centering
\includegraphics[width=0.78\textwidth]{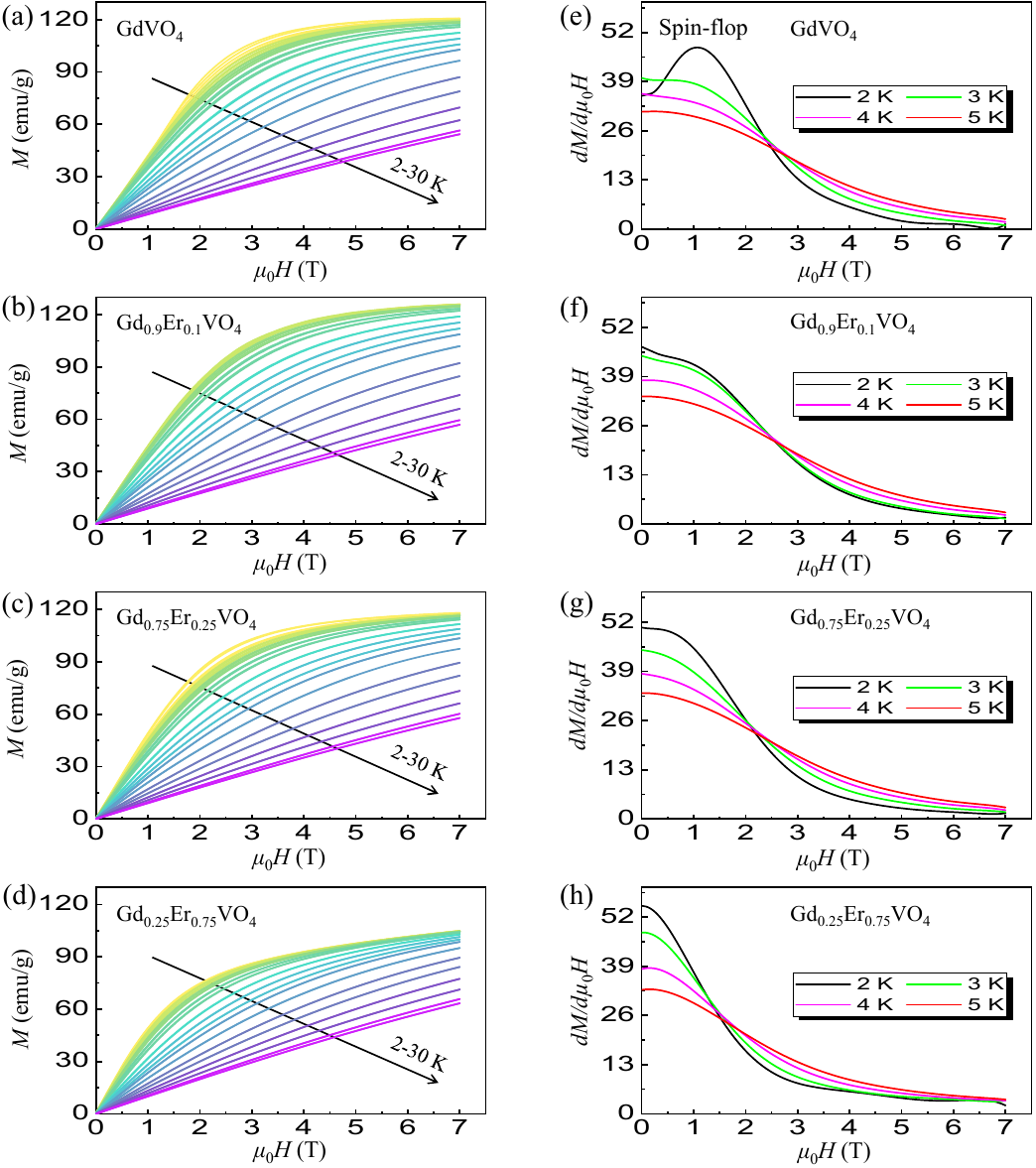}
\caption{
Magnetic-field dependence of the magnetization for Gd$_{1-x}$Er$_x$VO$_4$.
(a--d) Isothermal magnetization curves $M(\mu_0H)$ measured at low temperatures (2--30~K) with magnetic fields increasing from 0 to 7 T and then decreasing from 7 to 0 T. (e--h) Corresponding smoothed derivatives $dM/d(\mu_0H)$, highlighting the spin-flop-like field-induced anomaly. The anomaly observed in GdVO$_4$ becomes progressively weaker with increasing Er substitution. Raw data plots are provided in Fig. S1 of the Supplemental Material \cite{Suplemental-Raw-Data}.
}
\label{fig:Figure4}
\end{figure*}

Figure~\ref{fig:Figure4} presents isothermal magnetization curves measured between 2 and 30~K. No obvious magnetic hysteresis is observed, indicating that the magnetization processes are essentially reversible over the measured field range.

For GdVO$_4$, the magnetization increases nearly linearly at low fields and gradually approaches saturation at higher fields as the Gd$^{3+}$ moments align with the applied field. The derivative $dM/d(\mu_0H)$ exhibits a broad anomaly near $\mu_0H \approx 1$~T, consistent with a spin-flop-like field-induced anomaly \cite{li2016possible}. This feature indicates that the antiferromagnetic state in the parent compound lies close to a field-induced magnetic instability \cite{SpinFlop}.

\begin{figure*}[htbp]
\centering
\includegraphics[width=0.78\textwidth]{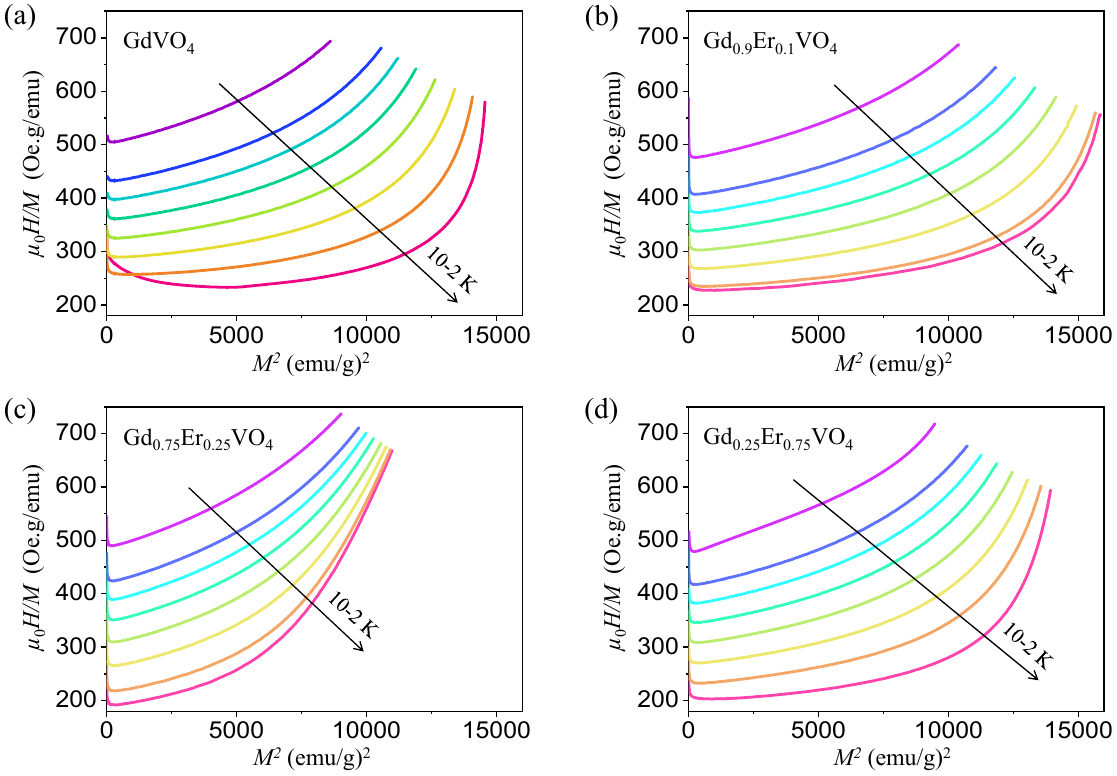}
\caption{
Arrott-type plots ($\mu_0H/M$ versus $M^2$) for Gd$_{1-x}$Er$_x$VO$_4$ measured between 2 and 10~K. (a) $x=0$, (b) $x=0.1$, (c) $x=0.25$, and (d) $x=0.75$. The negative slope observed for GdVO$_4$ at low temperature suggests a first-order-like character of the spin-flop-like field-induced anomaly.
}
\label{fig:Figure5}
\end{figure*}

\begin{figure*}[htbp]
\centering
\includegraphics[width=0.78\textwidth]{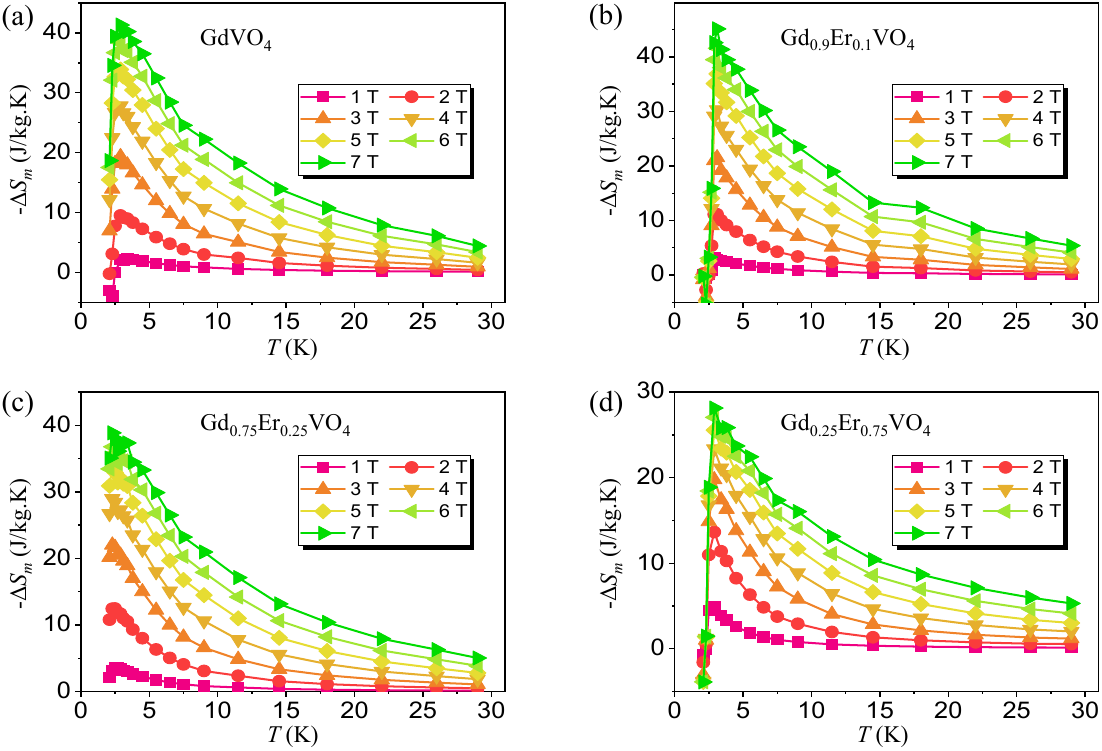}
\caption{
Temperature dependence of the magnetic entropy change $-\Delta S_{\mathrm{m}}$ for Gd$_{1-x}$Er$_x$VO$_4$ under different magnetic-field changes.
(a) $x=0$, (b) $x=0.1$, (c) $x=0.25$, and (d) $x=0.75$. The maximum entropy change occurs near the magnetic ordering temperature.
}
\label{fig:Figure6}
\end{figure*}

\begin{figure}[htbp]
\centering
\includegraphics[width=0.45\textwidth]{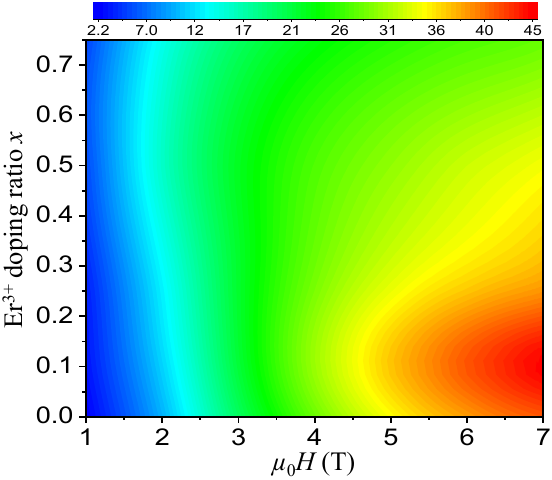}
\caption{
Contour maps of the magnetic entropy change $-\Delta S_{\mathrm{m}}(T,\mu_0H)$ for Gd$_{1-x}$Er$_x$VO$_4$. The color scale represents the magnitude of $|-\Delta S_{\mathrm{m}}|$ in $\mathrm{J\,kg^{-1}\,K^{-1}}$, illustrating the evolution of the magnetocaloric response with temperature and applied magnetic field.
}
\label{fig:Figure7}
\end{figure}

This anomaly evolves rapidly with Er substitution. As the Er content increases, the magnetization shows a reduced tendency toward saturation, and the anomaly in $dM/d(\mu_0H)$ becomes progressively weaker, eventually disappearing at higher Er concentrations. These results indicate that Er substitution substantially modifies the competition among exchange interactions, dipolar coupling, and magnetic anisotropy.

Arrott plots provide a complementary way to assess the character of the field-induced magnetic response. As shown in Fig.~\ref{fig:Figure5}(a), the low-temperature curve for GdVO$_4$ develops a negative slope at low $M^2$, suggesting a first-order-like character of the field-induced anomaly, whereas the Er-substituted samples remain positive over the full measured range. The field-induced magnetic response therefore evolves from a comparatively sharp instability in GdVO$_4$ to a broader and more gradual field-driven polarization process in the Er-substituted compositions.

\subsection{Magnetocaloric response and refrigeration performance}

The changes in magnetic ordering and field-induced magnetic response discussed above are directly reflected in the magnetocaloric properties. To quantify this connection, the magnetic entropy change was derived from the isothermal magnetization data using the Maxwell relation [Eq.~(\ref{eq:maxwell})]:
\begin{equation}
-\Delta S_{\mathrm{m}}(T,\mu_0H) = \int_0^{\mu_0H} \left(\frac{\partial M}{\partial T}\right)_{\mu_0H} d(\mu_0H).
\label{eq:maxwell}
\end{equation}

The resulting entropy changes are shown in Fig.~\ref{fig:Figure6}, and a clear compositional optimum is observed at low Er concentration. Among the investigated samples, Gd$_{0.9}$Er$_{0.1}$VO$_4$ exhibits the largest $-\Delta S_{\mathrm{m}}$, reaching its maximum near 3.1~K under $\mu_0 \Delta H = 7$~T. Contour plots of $-\Delta S_{\mathrm{m}}(T,\mu_0H)$ are shown in Fig.~\ref{fig:Figure7}, further illustrating the evolution of the entropy release with temperature and magnetic field.

A low Er concentration enhances the entropy change, whereas higher Er concentrations reduce it. This nonmonotonic behavior reflects the competition between two effects. On the one hand, weak Er substitution lowers $T_{\mathrm{N}}$ and shifts the entropy release toward the optimal low-temperature window while preserving much of the large Gd-derived spin entropy. On the other hand, at higher Er concentrations, the increasing role of magnetic anisotropy and the more gradual field-induced magnetic response reduce the field-driven entropy change.

\begin{figure}[t]
\centering
\includegraphics[width=0.45\textwidth]{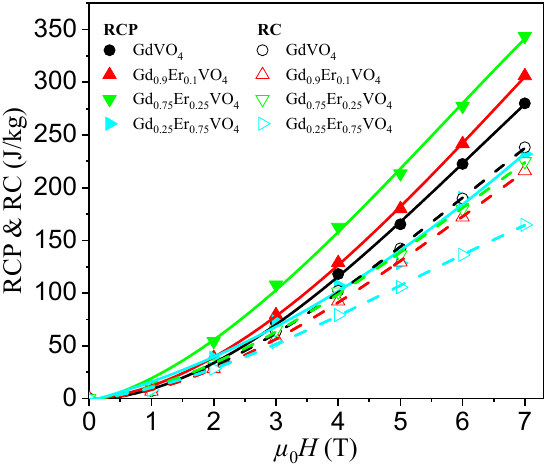}
\caption{
Relative cooling power (RCP) and refrigeration capacity (RC) of Gd$_{1-x}$Er$_x$VO$_4$ as functions of the applied magnetic field. Solid symbols represent RCP values and open symbols denote RC values. A low Er concentration enhances the cooling performance, whereas excessive Er content reduces both quantities.
}
\label{fig:Figure8}
\end{figure}

The cooling performance was further evaluated using the relative cooling power (RCP) and refrigeration capacity (RC), defined by Eqs.~(\ref{eq:rcp}) and (\ref{eq:rc}).
\begin{equation}
RCP = |-\Delta S_{\mathrm{m}}^{\max}| \times \delta T_{\mathrm{FWHM}},
\label{eq:rcp}
\end{equation}

\begin{equation}
RC = \int_{T_1}^{T_2} |\Delta S_{\mathrm{m}}(T,\mu_0H)|\, dT,
\label{eq:rc}
\end{equation}
where $\delta T_{\mathrm{FWHM}}$ is the full width at half maximum of the entropy-change peak. The results are summarized in Fig.~\ref{fig:Figure8}. Both quantities increase with magnetic field for all samples. Under a field change of 7~T, Gd$_{0.9}$Er$_{0.1}$VO$_4$ and Gd$_{0.75}$Er$_{0.25}$VO$_4$ exhibit improved RCP values compared with the end members.

To place the present compounds in context, Table~\ref{tab:table3} compares its magnetocaloric parameters with those of representative cryogenic refrigerants reported in the literature. For a fair comparison, the low-field magnetic entropy change is listed under a field change of 1~T whenever available; values reported at 2~T are explicitly marked. In addition to the experimental values of $-\Delta S_{\mathrm{m}}$, the table also includes the rare-earth mass fraction $M_R/M_W$, the theoretical magnetic entropy change, and the ion-normalized entropy change, which together provide a more intrinsic assessment of the magnetocaloric performance across different material families. The resulting ion-normalized values for GdVO$_4$ and Gd$_{0.9}$Er$_{0.1}$VO$_4$ are approximately 59.3 and 63.7, respectively, comparable to those of several previously reported oxide-based cryogenic refrigerants.

\begin{table*}[htbp]
\caption{\label{tab:table3}
Comparison of the magnetocaloric performance of the present compounds with representative cryogenic magnetic refrigerants reported in the literature.
Here $T_0$ denotes the characteristic magnetic transition temperature. For compounds without a well-defined long-range ordering temperature, the characteristic low-temperature anomaly reported in the literature is listed.
The column ``Low-field $-\Delta S_{\mathrm{m}}^{\mathrm{exp}}$'' lists the experimental magnetic entropy change under a magnetic-field change of 1~T unless otherwise noted; for the present work, the corresponding low-field values are given at 2~T.
$-\Delta S_{\mathrm{m}}^{\mathrm{exp}}(5~\mathrm{T})$ is the experimental magnetic entropy change under a magnetic-field change of 5~T.
All entropy-change values are given in units of J\,kg$^{-1}$\,K$^{-1}$.
$M_R/M_W$ denotes the mass fraction of rare-earth magnetic ions in the compound.
$-\Delta S_{\mathrm{m}}^{\mathrm{theory}}$ is the theoretical entropy change estimated from the full magnetic entropy of the corresponding magnetic ions.
The column ``Ion-normalized $-\Delta S_{\mathrm{m}}^{\mathrm{exp}}(5~\mathrm{T})$'' gives the magnetic-ion-normalized entropy change obtained by dividing $-\Delta S_{\mathrm{m}}^{\mathrm{exp}}(5~\mathrm{T})$ by $M_R/M_W$.}
\begin{ruledtabular}
\footnotesize
\setlength{\tabcolsep}{3pt}
\renewcommand{\arraystretch}{1.03}
\begin{tabular}{l
D{.}{.}{2.2}
c
D{.}{.}{2.1}
D{.}{.}{2.1}
D{.}{.}{2.1}
c
c}
Compound &
\multicolumn{1}{c}{$T_0$} &
\multicolumn{1}{c}{Low-field$^\ddag$ $-\Delta S_{\mathrm{m}}^{\mathrm{exp}}$} &
\multicolumn{1}{c}{$-\Delta S_{\mathrm{m}}^{\mathrm{exp}}(5~\mathrm{T})$} &
\multicolumn{1}{c}{$M_R/M_W$} &
\multicolumn{1}{c}{$-\Delta S_{\mathrm{m}}^{\mathrm{theory}}$} &
\multicolumn{1}{c}{Ion-normalized $-\Delta S_{\mathrm{m}}^{\mathrm{exp}}(5~\mathrm{T})$} &
Ref. \\
&
\multicolumn{1}{c}{(K)} &
\multicolumn{1}{c}{(J\,kg$^{-1}$\,K$^{-1}$)} &
\multicolumn{1}{c}{(J\,kg$^{-1}$\,K$^{-1}$)} &
\multicolumn{1}{c}{(\%)} &
\multicolumn{1}{c}{(J\,kg$^{-1}$\,K$^{-1}$)} &
\multicolumn{1}{c}{(J\,kg$^{-1}$\,K$^{-1}$)} &
\\
\hline

\multicolumn{8}{l}{\textit{This work} (low-field data at 2~T)} \\

\textbf{GdVO$_4$} & 3.65 & 9.5$^\dag$ & 34.3 & 57.8 & 63.6 & 59.3 & This work \\
\textbf{Gd$_{0.9}$Er$_{0.1}$VO$_4$} & 2.76 & 11.1$^\dag$ & 36.9 & 57.9 & 62.0 & 63.7 & This work \\
\textbf{Gd$_{0.75}$Er$_{0.25}$VO$_4$} & \multicolumn{1}{c}{$<2$} & 12.5$^\dag$ & 33.7 & 58.1 & 59.5 & 58.0 & This work \\
\textbf{Gd$_{0.5}$Er$_{0.5}$VO$_4$} & \multicolumn{1}{c}{$<2$} & 15.3$^\dag$ & 29.9 & 58.5 & 55.5 & 51.1 & This work \\
\textbf{Gd$_{0.25}$Er$_{0.75}$VO$_4$} & \multicolumn{1}{c}{$<2$} & 13.6$^\dag$ & 25.6 & 58.9 & 51.4 & 43.5 & This work \\

\hline

\multicolumn{8}{l}{\textit{Rare-earth halides}} \\

EuCl$_2$ & 1.69 & 36.8 & 74.6 & 68.2 & 77.8 & 109.4 & \cite{EuCl2} \\
CsEuCl$_3$ & 1.1 & 8.2 & 38.6 & 38.9 & 44.2 & 99.2 & \cite{CsEuCl3JAP2025} \\

\hline

\multicolumn{8}{l}{\textit{Rare-earth fluorides and hydroxides}} \\

Gd(OH)F$_2$ & 0.5 & 21.3 & 68.7 & 74.1 & 81.5 & 92.7 & \cite{GdOHF2} \\
NH$_4$GdF$_4$ & 0.85 & 38.2 & 64.9 & 62.6 & 68.8 & 103.7 & \cite{NH4GdF4} \\

\hline

\multicolumn{8}{l}{\textit{Rare-earth oxides and oxysalts}} \\

EuB$_4$O$_7$ & \multicolumn{1}{c}{$<0.4$} & 22.8 & 47.6 & 49.5 & 56.3 & 96.2 & \cite{EuB4O7} \\
Eu$_2$SiO$_4$ & 5.8 & 21.6 & 52.8 & 76.7 & 87.2 & 68.9 & \cite{SJ2025} \\
Gd$_3$Ga$_5$O$_{12}$ & 0.8 & 21$^\dag$ & 35.0 & 46.6 & 51.2 & 75.1 & \cite{Gd3Ga5O12} \\
GdCrO$_3$ & 6.74 & 3.1 & 37.0 & 61.1 & 67.2 & 60.6 & \cite{GdCrO3} \\
Gd$_2$WO$_6$ & \multicolumn{1}{c}{$<2$} & 2.9 & 25.2 & 52.9 & 58.2 & 47.6 & \cite{Gd2WO6APL2025} \\
GdAlO$_3$ & 3.9 & 0.5 & 23.6 & 67.7 & 74.5 & 34.8 & \cite{GdAlO3} \\
GdBO$_3$ & \multicolumn{1}{c}{0.61, 1.72} & 6.8$^\dag$ & 35.6 & 72.8 & 80.1 & 48.9 & \cite{LnBO3} \\
Gd$_2$B$_2$MoO$_9$ & 0.8 & 13.3 & 55.0 & 54.6 & 60.0 & 100.7 & \cite{LLW2026NC} \\

\hline

\multicolumn{8}{l}{\textit{Cluster and molecular refrigerants}} \\

Gd$_{152}$Ni$_{14}$@Cl$_{24}$ & \multicolumn{1}{c}{$<2.5$} & 8.1 & 46.1 & 57.6 & 66.4 & 80.0 & \cite{Gd152Ni14} \\
Gd$_{60}$ & \multicolumn{1}{c}{$<2$} & 4.1 & 39.1 & 47.5 & 52.2 & 82.3 & \cite{Gd60} \\

\end{tabular}
\end{ruledtabular}
\begin{flushleft}
$^\ddag$Low-field values correspond to a magnetic-field change of 1~T unless otherwise noted. \\
$^\dag$Values reported for a magnetic-field change of 2~T.
\end{flushleft}
\end{table*} 

It is also noteworthy that Er substitution enhances the low-field magnetocaloric response. As summarized in Table~\ref{tab:table3}, the experimental magnetic entropy change under a field change of 2~T increases from 9.5~J\,kg$^{-1}$\,K$^{-1}$ for GdVO$_4$ to 15.3~J\,kg$^{-1}$\,K$^{-1}$ for Gd$_{0.5}$Er$_{0.5}$VO$_4$, and then decreases slightly to 13.6~J\,kg$^{-1}$\,K$^{-1}$ for Gd$_{0.25}$Er$_{0.75}$VO$_4$. This behavior indicates that Er substitution not only suppresses the magnetic ordering temperature of GdVO$_4$, but also improves the low-field magnetocaloric performance of the substituted compounds. Notably, the 2~T entropy change of the Er-substituted samples approaches the value reported for single-crystalline Gd$_3$Ga$_5$O$_{12}$, underscoring the effectiveness of chemical substitution in improving the cryogenic magnetocaloric response of zircon-type rare-earth vanadates. Although the low-field entropy change reaches its maximum at intermediate Er concentrations, the overall magnetocaloric performance under higher fields remains optimal for Gd$_{0.9}$Er$_{0.1}$VO$_4$, owing to its more favorable balance between suppressed magnetic ordering and retained Gd-derived spin entropy.

Overall, the results indicate that weak Er substitution effectively tunes the low-temperature magnetic response and enhances the magnetocaloric performance. In particular, Gd$_{0.9}$Er$_{0.1}$VO$_4$ exhibits a favorable combination of entropy change and cooling capacity, consistent with an optimized balance between suppressed magnetic ordering and retained rare-earth spin entropy.

\section{Conclusion}

In summary, we have investigated the structural, magnetic, and magnetocaloric properties of polycrystalline Gd$_{1-x}$Er$_x$VO$_4$ ($x = 0$, $0.1$, $0.25$, $0.5$, $0.75$). Future studies could focus on exploring the effects of different rare-earth substitutions on the magnetocaloric properties of zircon-type vanadates to further enhance their performance. All compositions retain the zircon-type tetragonal structure, while Er substitution induces a systematic lattice contraction and modifies the local environment of the rare-earth sublattice. These structural changes are accompanied by a clear evolution of the magnetic behavior. In particular, the N{\'e}el temperature is suppressed with Er substitution, and the spin-flop-like field-induced anomaly observed in GdVO$_4$ becomes progressively weaker, indicating that Er substitution modifies the competition among exchange interactions, dipolar coupling, and magnetic anisotropy.

This evolution directly impacts the magnetocaloric performance. A low Er concentration improves the low-temperature MCE, with Gd$_{0.9}$Er$_{0.1}$VO$_4$ exhibiting the best performance among the investigated compositions, including a maximum magnetic entropy change of 45.1~$\mathrm{J\,kg^{-1}\,K^{-1}}$ for $\mu_0 \Delta H = 7$~T and 36.9~$\mathrm{J\,kg^{-1}\,K^{-1}}$ for $\mu_0 \Delta H = 5$~T. The corresponding magnetic-ion--normalized entropy change further supports an enhancement of the intrinsic refrigerant efficiency at low Er concentration.

More broadly, the present results show that weak rare-earth substitution can be used to tune the competition between magnetic ordering and available spin entropy in zircon-type vanadates. Gd$_{1-x}$Er$_x$VO$_4$ therefore provides a useful platform for exploring how chemical pressure and anisotropy cooperate in determining cryogenic magnetocaloric behavior in rare-earth oxides.

\section{Acknowledgments}

This work was supported by the Science and Technology Development Fund of the Macao S.A.R. (File Nos. 0104/2024/AFJ, 0115/2024/RIB2, and 0002/2024/TFP), the University of Macau (MYRG-GRG2024-00158-IAPME and MYRG-GRG2025-00251-IAPME), and the Guangdong-Hong Kong-Macao Joint Laboratory for Neutron Scattering Science and Technology (Grant No. 2019B121205003).

\section{Data Availability}

The data that support the findings of this article are not publicly available. The data are available from the authors upon reasonable request.

\vfill

\bibliography{GdVO4}

\end{document}